\documentclass[manuscript, nonacm]{acmart}

\usepackage{subfigure}
\usepackage{rotating}
\AtBeginDocument{%
  }

\begin{document}

\title{Navigating the State of Cognitive Flow: Context-Aware AI Interventions for Effective Reasoning Support}

\begingroup
\renewcommand\thefootnote{}\footnote{
 This paper was presented at the 2025 ACM Workshop on Human-AI Interaction for Augmented Reasoning (AIREASONING-2025-01). This is the authors’ version for arXiv.}
\endgroup

% \author{Anonymous Authors}/
\author{Dinithi Dissanayake}
\email{dinithi@ahlab.org}
\orcid{0009-0007-0178-984X}
\affiliation{%
  \institution{Augmented Human Lab, National University of Singapore}
  \country{Singapore}
}
% \affiliation{%
\author{Suranga Nanayakkara}
\email{suranga@ahlab.org}
\affiliation{%
  \institution{Augmented Human Lab, National University of Singapore}
  % \city{National University of Singapore}
  \country{Singapore}
}

% \affiliation{%

% \affiliation{%

% \affiliation{%

% \affiliation{%

% \affiliation{%

% \affiliation{%

% \affiliation{%


\begin{abstract}
Flow Theory describes an optimal cognitive state where individuals experience deep focus and intrinsic motivation when a task’s difficulty aligns with their skill level. In AI-augmented reasoning, interventions that disrupt the state of cognitive flow can hinder rather than enhance decision-making. This paper proposes a context-aware cognitive augmentation framework that adapts interventions based on three key contextual factors: type, timing, and scale. By leveraging multimodal behavioral cues (e.g., gaze behavior, typing hesitation, interaction speed), AI can dynamically adjust cognitive support to maintain or restore flow. We introduce the concept of cognitive flow, an extension of flow theory in AI-augmented reasoning, where interventions are personalized, adaptive, and minimally intrusive. By shifting from static interventions to context-aware augmentation, our approach ensures that AI systems support deep engagement in complex decision-making and reasoning without disrupting cognitive immersion.
\end{abstract}

\maketitle

\section{Introduction}

The theory of flow, proposed by Csikszentmihalyi \cite{csikszentmihalyi1990flow}, defines an optimal psychological state in which individuals experience deep focus and intrinsic motivation when the challenge of a task is perfectly matched to their skill level. When a task is too easy, users become bored, while excessive difficulty leads to frustration. For example, in video games, the enjoyment of a game depends on several factors, including the player’s skill level and the challenge the game presents. A game that appropriately matches a player’s skill level fosters a sense of engagement, encouraging continued play. Similarly, any new intervention or augmentation should support the user in either attaining or maintaining the state of flow of whatever the current task the user is engaged in.

In the domain of AI-augmented reasoning systems, the goal is to enhance human decision-making by providing intelligent feedback on logical reasoning, bias detection, and argumentation \cite{danry2023don, danry2020wearable, le2018cognitive}. The effectiveness of such systems depends on several factors, with the type, timing, and scale of the intervention playing crucial roles in determining the quality of the augmentation. For example, poorly timed or intrusive interventions can disrupt a user’s state of cognitive flow (whereby we define the optimal cognitive state one would prefer), leading to disengagement rather than improving reasoning.

Current AI reasoning assistants often adopt a static or one-size-fits-all approach, assuming that interventions should be uniformly applied across all users \cite{danry2020wearable}. However, individuals respond to cognitive support in diverse ways \cite{tanprasert2024debate}. Some may prefer direct interventions, such as explicit fact-checking and counterarguments, while others may benefit from Socratic questioning, which promotes self-guided reflection. The timing of the intervention is equally significant—interrupting a user mid-thought with a suggestion can break their focus, whereas the ability to understand subtle nudges, such as gaze-based or gesture-based cues, might be able to guide the intervention to present their reasoning seamlessly without causing disruption. Thereby, we identify three key factors that impact the effectiveness of these interventions:
\begin{itemize} \item \textbf{Type of intervention}: Some individuals may prefer direct feedback, while others may respond better to Socratic questioning.
\item \textbf{Timing of intervention}: Understanding when users are "stuck" through subtle cues, such as gaze or gesture-based signals, is critical.
\item \textbf{Scale of intervention}: The magnitude of the intervention—whether subtle or more direct—also determines its impact.
\end{itemize}

These three factors; type, timing, and scale, determine the effectiveness of interventions, which can be broadly categorized under the concept of \textit{context}. This paper proposes that context-aware reasoning interventions should be designed to dynamically adapt based on real-time user engagement signals, ensuring that interventions enhance rather than disrupt the cognitive flow. Contextual awareness plays a pivotal role in identifying the appropriate type, timing, and scale of interventions, thus enabling personalized cognitive assistance. By maintaining this balance, the intervention fosters a sense of autonomy in task completion, ensuring that users feel they have accomplished the task themselves, rather than perceiving it as something imposed by an external system, thus supporting the notion of assistive augmentation \cite{tan2025assistive}.

With recent developments in multimodal AI models \cite{dissanayake2025vrsense, Johann2023CHI, acosta2022multimodal}, deriving contextual awareness becomes a matter of mapping input modalities to cognitive assistance and understanding when, what, and how to intervene. By leveraging multimodal cues, such as gaze behavior, typing hesitation, and interaction patterns, AI systems can infer cognitive load and deliver interventions at moments that sustain deep reasoning rather than hinder it. Furthermore, longitudinal engagement data can be utilized to cluster users based on preferred reasoning styles and cognitive engagement patterns, enabling AI systems to personalize augmentation strategies more effectively.

By grounding AI-augmented reasoning systems in Flow Theory, we can shift from rigid, one-size-fits-all intervention models to truly adaptive cognitive assistants that support reasoning in a natural, personalized, and minimally disruptive way. This approach has the potential to enhance critical thinking without breaking immersion, making AI-augmented reasoning a more intuitive and effective tool for human cognition.

\section{State of Flow --> State of Cognitive Flow}

\begin{figure}[h]
  \centering
  \includegraphics[width=0.75\linewidth]{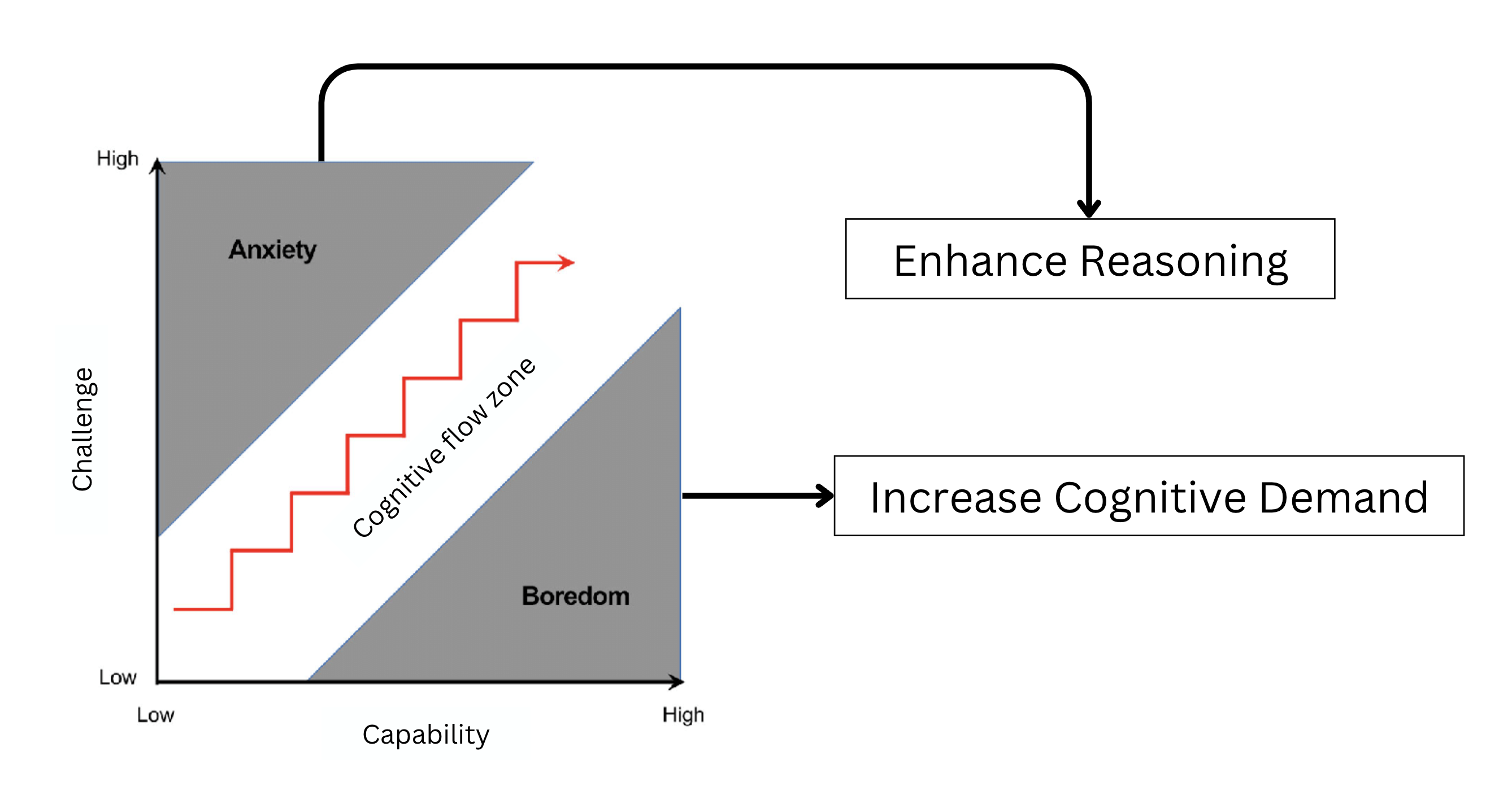}
  \caption{State of Cognitive Flow}
  \label{fig: genrevsms}
\end{figure}

One of the fundamental questions in positive psychology concerns the nature of a \textit{good life}. Flow theory offers one perspective, emphasizing \textit{deep engagement and full immersion} in an activity as a key component of well-being. Csikszentmihalyi \cite{csikszentmihalyi1990flow,nakamura2009flow} defines \textit{flow} as a psychological state in which individuals experience complete absorption in a task, striking an optimal balance between challenge and skill. When this balance is achieved, individuals remain fully engaged, motivated, and focused on the task at hand.  

Because flow is shaped by both personal and environmental factors, it represents an emergent motivational process \cite{csikszentmihalyi1990flow}. That is the goals and motivations driving an individual’s engagement evolve dynamically in response to immediate situational factors rather than being dictated solely by predefined traits, roles, or scripts. This dynamic nature is particularly relevant in the context of AI-augmented cognitive interventions, where external assistance must be carefully calibrated to avoid disrupting or undermining an individual’s sense of agency and engagement, thereby positing the importance of \textit{contextual awareness} for these external cognitive augmentation systems.  

For cognitive augmentation to be effective, it must support rather than disrupt the state of cognitive flow. The goal is to ensure that reasoning tasks remain appropriately challenging, not so difficult as to cause frustration, nor so easy as to induce disengagement. AI-driven interventions can play a pivotal role in helping individuals reach or maintain cognitive flow by:

\begin{itemize}
    \item \textbf{Enhancing reasoning when challenges exceed an individual's current capabilities}, providing just enough support to facilitate deeper engagement.  
    \item \textbf{Increasing cognitive demand when tasks are too easy}, introducing counterarguments, critiques, or prompts that encourage deeper critical thinking.  
\end{itemize}

Thus, we define the optimal cognitive state as \textbf{cognitive flow} (Figure \ref{fig: genrevsms}), wherein AI-augmented reasoning systems dynamically adjust interventions to either challenge or sustain the user’s engagement level. By maintaining this balance, AI can foster a sense of agency in decision-making, ensuring that individuals feel they have arrived at conclusions through their own reasoning processes rather than external imposition. This approach aligns with the broader goal of empowering individuals to engage in more deliberate, reflective, and well-reasoned thought processes.

\section{Bridging AI-Driven Behavioral Modeling with Cognitive Flow}
Recent advances in AI and the increasing availability of large-scale multimodal datasets have made it possible to model human behavior with unprecedented granularity. By leveraging behavioral cues—such as gaze anticipation, gesture patterns, typing patterns, and physiological signals—AI systems can infer user states and dynamically adapt cognitive interventions \cite{cha2022correlation,zimmerer2022reducing}. This enables a shift from static, one-size-fits-all interventions to context-aware augmentation strategies tailored to an individual's cognitive state. In the context of cognitive flow, AI systems must intelligently map multimodal inputs to determine whether an individual is in a state of deep engagement, facing cognitive overload, or experiencing disengagement due to insufficient challenge. Learned representations derived from behavioral data provide a means to infer cognitive load and engagement levels, enabling real-time adjustments to interventions.
% 
% One other promising approach is to cluster users based on longitudinal behavioral patterns, allowing AI-driven augmentation to personalize interventions over time. By observing how users respond to different levels and types of augmentation, AI systems can infer policies governing when and how users shift between cognitive states. This enables:

% Personalized Intervention Strategies: Users with similar cognitive adaptation patterns can be grouped, allowing for tailored augmentation that optimally sustains flow.
% Dynamic Context Awareness: AI can predict shifts in cognitive states based on past data, refining intervention strategies as users transition between tasks and cognitive demands.
% Adaptive Flow Regulation: 
By balancing challenge and support based on real-time multimodal inputs, the system ensures that users neither disengage due to excessive difficulty nor lose interest due to low complexity.
By integrating context-aware AI-driven augmentation with flow theory, we move toward systems that do more than assist with reasoning—they optimize the conditions under which reasoning and critical thinking can thrive. Rather than disrupting engagement, AI enhances cognition in ways that feel seamless and natural, preserving a sense of agency while fostering deeper intellectual engagement.

\section{Call to Action}
This paper presents a case for the importance of context awareness in AI-driven cognitive interventions. We introduce the concept of Cognitive Flow Alignment, emphasizing that interventions are most effective when they dynamically adjust to an individual's cognitive state, neither disrupting engagement nor allowing stagnation. Using insights from flow theory, we argue that AI systems should infer behavioral cues from multimodal data to sustain an optimal cognitive state. While current AI-based augmentation focuses on task performance and engagement metrics, we propose that future systems should be evaluated on their ability to contextually regulate cognitive flow. This requires a shift from static intervention strategies to adaptive models that personalize intervention type, intensity, and timing based on real-time user states. 
% Additionally, as AI learns longitudinal patterns of cognitive adaptation, new evaluation benchmarks must capture not just accuracy but also ecological validity—ensuring that interventions feel intuitive, enhance user agency, and align with natural cognitive rhythms. 
We advocate for a mixed-method evaluation framework combining behavioral analytics with subjective feedback to refine personalized intervention strategies that optimize cognitive augmentation while respecting individual differences in engagement and cognitive load.

\bibliographystyle{ACM-Reference-Format}
\bibliography{authordraft}

% %%
% %% If your work has an appendix, this is the place to put it.
%\appendix

%\input{Hopu_Appendix}









\end{document}